# Acoustojet: acoustic analogue of photonic jet phenomenon

Igor V. Minin[*], Oleg V. Minin

[1]*Tomsk State University, Lenina Ave. 36, Tomsk 634050, Russia*
*Corresponding author: prof.minin@gmail.com*

It has been demonstrated for the first time that an existence of acoustic analogue of photonic jet phenomenon, called acoustojet, providing for subwavelength localization of acoustic field in shadow area of arbitrary 3D penetrable mesoscale particle, is possible.

doi:10.3788/COLXXXXXX.XXXXXX.

The ability to control wave propagation is of fundamental interest in many areas of physics. But using conventional lenses, it's not possible to focus sound waves to a spot size smaller than half the wavelength of the radiation due to diffraction. Currently, different groups of researches make attempts to overcome diffraction limit. Overcoming diffraction limit means focusing radiation into the spot with sizes less than Airy disk [1]. In acoustic some new ideas (in contrast to lenses from conventional [2] and artificial [3]) materials were considered at last time: a linear array of closely spaced sound transducers was offered to produce a subwavelength-focused intensity profile at a distance of a quarter wavelength [4], to shape the flow of waves at subwavelength scales the metamaterials were investigated at [5-6], a broadband sounds can be also focused on a subwavelength scale using acoustic resonators [7], acoustic GRIN lens [8], artificial impedance surface with spatially varying values [9] etc.

However, existing methods of subwavelength focusing in acoustics as a rule based on new artificial materials, or require complex signal processing methods.

A short time ago an existence of "photonic nanojet" (PNJ) effect was paid attention on for the first time in 2004 [10], during investigation of laser radiation scattering on transparent quartz microcylinders and later on spherical low loses dielectric particles. It was demonstrated that, when plane wave fell at spherical particle, spatial resolution could reach up to one third of wavelength, which is lower than classical diffraction limit [11]. The review of current state on photonic jet formation by arbitrary shaped dielectric particles in electromagnetic spectrum is given in the work [12].

At the same time in acoustics and ultrasound there is much tension around the issue of possibility of sub-wave focusing of acoustic and ultrasound fields. In the ref. [13] a problem of plane acoustic wave focusing by spherical cavity (lens) filled with various gases was discussed. The Mie parameter of spherical particle was $\frac{2\pi R}{\lambda}$ = (17.5 ... 27.5). It was noted [13] that with small Mie parameters the location of focus near the shadow surface of spherical lens did not correspond to focal distance, prognosticated by geometrical optics theory.

On the other hand, in the linear mode it should be expected that methods of sub-wave focusing, based on photonic jet effect, can be successfully used in the acoustic range as well. Formally this can be stated based on the analogy between the equations, specifying acoustic and electromagnetic wave processes (see Table 1) [14].

Table 1
Analogy between acoustic and electromagnetic parameters and material characteristics [14]

| Acoustic parameters | Electromagnetic parameters | Analogy |
| --- | --- | --- |
| Acoustic pressure, P | Magnetic field, $H_z$ | $p \equiv H_z$ |
| Acoustic velocity, v | Electric field, E | $v_x \equiv E_y$ and $v_y \equiv -E_x$ |
| Dynamic density, $\rho_e(\omega)$ | Complex permittivity, $\varepsilon' = \varepsilon - j\sigma/\omega$ | $\rho_e \equiv \varepsilon$ |
| $K(\omega)$ dynamic volume modulus of elasticity | Magnetic permeability, μ | $K \equiv 1/\mu$ |

To shown the possibilities of acoustic analogue of photonic jet existences we examine multiply acoustic wave scattering at penetrable spherical particle. The problem of scattering on homogeneous limited area, placed inside the other homogeneous area, in the first approximation in the harmonic case, can be solved based on the Helmholtz equation [15].

Preliminary simulation results of scalar acoustic wave scattering at penetrable sphere are as follows. The acoustic contrast (the relation of wave numbers or sound velocity in the sphere material to the medium) C = 1.1, relative density (sphere material density to medium density) in all examples below is 1.2, sphere radius is given in the units of wavelength in medium.

Simulations shown that dependences of maximal field intensity along the jet vs spherical particle radius is almost linear and changes from about 10 for particle with radius 1 to 85 for particle with radius of 5 wavelength.

In Table 2 the summary characteristics of forming acoustic jets from contrast of sound velocities in sphere and medium material are given, where Lz – the length of acoustic jet at Full Width at Half Maximum (FWHM), x – the acoustic beam waist at FWHM in the point of maximal field intensity I along a jet. The values of Lz and x are given in the unit of wavelength in medium.

These results of modeling of acoustic plane wave scattering on penetrable sphere demonstrate that in the acoustic band a formation of acoustic analogue of photonic jet with subwavelength size is possible. Thus, as follows form the Table 2, with the parameter C=1.2 the half-width of acoustic field intensity distribution in the jet area makes up 0.42 of wavelength, and with C=1.4 it comprises 0.28 of wavelength, which agrees with analogous values for photonic jets of electromagnetic range by the order of values [10,11].

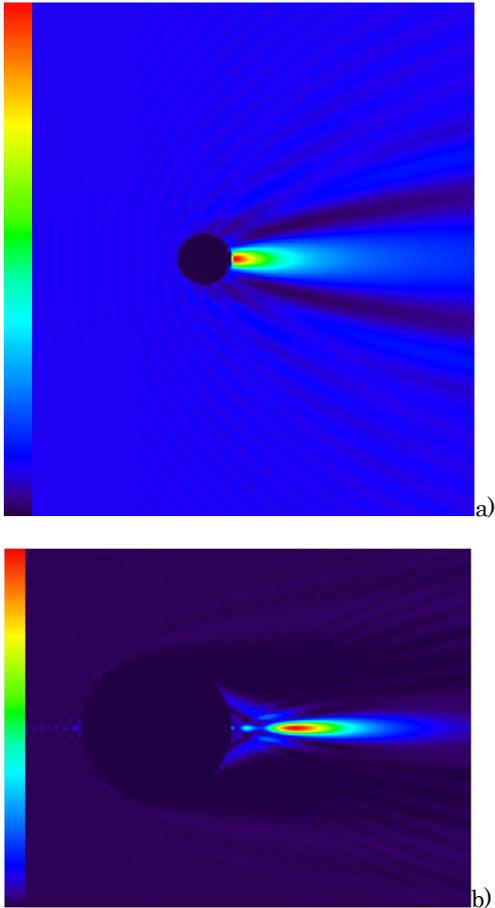

Fig.1. Scattering on penetrable spheres with different diameters: R=1 (a) and R=6 (b). Not in scale.

Table 2
Main parameters of acoustic jets

| C | Lz | x | I |
|---|---|---|---|
| 0.9 | 2 | 0.56 | 2 |
| 1.05 | 5.2 | 1.12 | 4.2 |
| 1.1 | 3.2 | 0.56 | 10.5 |
| 1.2 | 1.4 | 0.42 | 32 |
| 1.4 | 0.4 | 0.28 | 95 |

These results of modeling of acoustic plane wave scattering on penetrable sphere demonstrate that in the acoustic band a formation of acoustic analogue of photonic jet with subwavelength size is possible.

As practical example, formation of acoustojet for spherical particle (radius 5 of wavelengths) from lead placed in water is shown in fig. 2. Formation of localized sub-wave zone of acoustic field in the conditions of significant reflection, caused by large value of sphere material density compared to environment, is clearly visible. Maximal intensity in acoustic jet is over 60 times higher than intensity of incident field. Experimental verification of acoustojet will be conducted in near future.

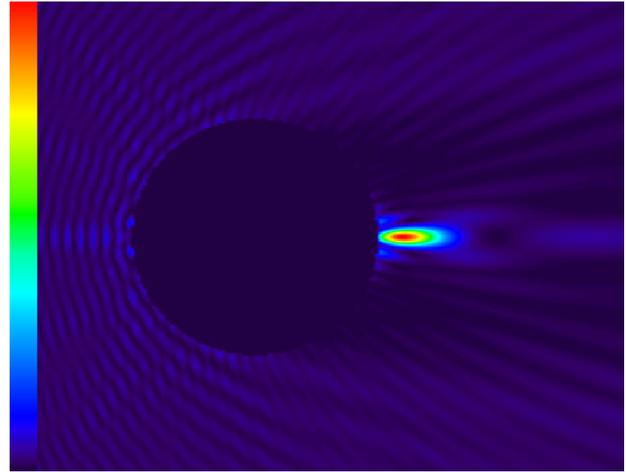

Fig.2. Formation of acoustojet from lead sphere particle in water: relative density contrast is 11,34, velocity of sound contrast is 1,44, radius of sphere is 5 wavelength in water.

There is reason to believe that discovered effect will be right for penetrable particles of arbitrary 3D shape as well, including also the reflection mode [12]. Application of the subwavelength acoustojet may be found in such areas as a miniature mesoscale acoustic lens, unique devices for microstructure examination and nondestructive testing based on acoustic microscopes [16,17], using which a microstructure of almost any optically opaque object can be investigated, acoustic traps of micro-particles [18], biological sensors [19,20], nondestructive testing systems [21] and so on.

Acknowledgment
This work was partially supported by the Mendeleev scientific fund of Tomsk State University № 8.2.48.2015